\documentclass[conference]{IEEEtran}  
\usepackage{graphicx}
\usepackage{color}
\usepackage{epsfig}
\usepackage{amsmath}
\usepackage{amssymb, amsfonts}
\usepackage{latexsym}
\newcommand{\boldgamma}{\mbox{\boldmath{$\gamma$}}}
\newcommand{\mean}[1]{\langle #1 \rangle}
\newcommand{\expectation}[2]{\mathbb{E}_{#2}\left[#1\right]}
\newcommand{\IX}{I_{\mathcal X}}
\newcommand{\popt}{p^{\rm opt}}
\newcommand{\lp}{\wp}
\newcommand{\boldlp}{\mbox{\boldmath{$\lp$}}}
\newcommand{\plt}{\mathbf{p}_{\rm lt}}
\newcommand{\pltopt}{\mathbf{\popt_{\rm lt}}}

\newcommand{\ptw}{p^{\rm tw}}

\newcommand{\pref}{p^{\rm ref}}

\newcommand{\sstar}{s^{\star}}
\newcommand{\wstar}{w^{\star}}

\newcommand{\MMSE}{{\rm MMSE}_{\mathcal X}}
\newcommand{\Pout}{P_{\rm out}}

\newcommand{\ptwfill}{p^{\rm tw}}
\newcommand{\openone}{\leavevmode\hbox{\small1\normalsize\kern-.33em1}}
\newtheorem{theorem}{Theorem}
\newtheorem{lemma}{Lemma}

\title{Power Allocation for Discrete-Input Non-Ergodic Block-Fading Channels}

\author{
\authorblockN{Khoa D. Nguyen}
\authorblockA{Institute for Telecommunications Research\\
University of South Australia \\
dangkhoa.nguyen@postgrads.unisa.edu.au}
\and
\authorblockN{Albert Guill{\'e}n i F{\`a}bregas}
\authorblockA{Engineering Department \\
University of Cambridge\\
guillen@ieee.org}
\and
\authorblockN{Lars K. Rasmussen}
\authorblockA{Institute for Telecommunications Research\\
University of South Australia \\
lars.rasmussen@unisa.edu.au}
}

\date{}
\begin{document}

\maketitle

\footnotetext[1]{This work has been partly supported by the Australian Research Council under ARC grants RN0459498 and DP0558861.}

\begin{abstract}
We consider power allocation algorithms for fixed-rate transmission over
Nakagami-$m$ non-ergodic block-fading channels with perfect transmitter and
receiver channel state information and discrete input signal constellations
under short- and long-term power constraints. Optimal power allocation
schemes are shown to be direct applications of previous results in the
literature. We show that the SNR exponent of the optimal short-term scheme is
given by the Singleton bound. We also illustrate the significant gains
available by employing long-term power constraints. Due to the nature of the
expressions involved, the complexity of optimal schemes may be prohibitive for
system implementation. We propose simple sub-optimal power allocation schemes
whose outage probability performance is very close to the minimum outage
probability obtained by optimal schemes.
\end{abstract}

\section{Introduction}

The non-ergodic block-fading channel introduced in \cite{OzarowShamaiWyner1994}
and \cite{BiglieriProakisShamai1998} models communication scenarios where each
codeword spans a fixed number of independently faded blocks. The block-fading
channel is an accurate model for slowly-varying fading scenarios
encountered with slow time-frequency hopping or orthogonal frequency division
multiplexing (OFDM).

Since each codeword experiences a finite number of degrees of freedom, the channel is
non-ergodic. Therefore, the channel has zero capacity under most common
fading statistics. A useful measure for the channel reliability in non-ergodic
channels is the outage probability, which is the probability that a given
communication rate is not supported by the channel
\cite{OzarowShamaiWyner1994,BiglieriProakisShamai1998}. The outage probability
is the lowest possible word error probability for sufficiently long codes.

When knowledge of channel parameters, referred to as channel state information
(CSI), is not available at the transmitter, transmit power is allocated
uniformly over the blocks.
When CSI is available at the transmitter, power allocation techniques can be
used to increase the instantaneous mutual information, thus improving the
outage performance. Optimal power allocation schemes, minimizing the outage
probability, have been studied under various power constraints. For systems
with short-term power constraints (per-codeword power constraint),
water-filling is the optimal power allocation scheme \cite{CoverThomas2006}. In
\cite{CaireTariccoBiglieri1999} the power allocation problem is solved under
long-term power constraint, showing that remarkable gains are possible with
respect to short-term power allocation. For channels with two or more fading
blocks, zero outage can be obtained under long-term power constraint. In both
cases, the optimal input distribution is Gaussian.

In \cite{LozanoTulinoVerdu2006}, the authors propose the optimal short-term
power allocation scheme to maximize the mutual information of parallel channels
for arbitrary input distributions. Also, as mentioned in
\cite{LozanoTulinoVerdu2006}, optimal short-term power allocation for
block-fading channels with discrete inputs is obtained directly from their
results. Due to its complexity, the optimal solution in
\cite{LozanoTulinoVerdu2006} does not provide much insight into the impact of the
parameters involved, and may also prohibit the application to systems with
strict memory and computational constraints.

In this paper, we study optimal short- and long-term power allocation schemes for {\em fixed-rate} transmission over discrete-input block-fading channels with perfect CSI at the transmitter and receiver. We consider {\em non-causal} CSI, namely, the channel gains corresponding to the transmission of one codeword are known to the transmitter and receiver. In practice, this non-causal assumption reflects the situation of OFDM, where the time-domain channel is estimated but the signals are transmitted in the frequency domain.
In particular, we show that the SNR exponent for optimal short-term allocation
is given by the Singleton bound
\cite{KnoppHumblet2000,MalkamakiLeib1999,FabregasCaire2006}. Furthermore, we
show that the results in \cite{LozanoTulinoVerdu2006} are instrumental in
obtaining the optimal long-term solution. We further aim at reducing the
complexity drawbacks of optimal schemes by proposing suboptimal short- and
long-term power allocation schemes. The suboptimal schemes are simpler as
compared to the corresponding optimal schemes, yet they suffer only negligible
losses compared to the optimal performance. Proofs of all results can be found in \cite{NguyenGuillenRasmussen2007}.

\section{System Model}
\label{se:model}

Consider transmission over an additive white Gaussian noise (AWGN) block-fading
channel with $B$ blocks of $L$ channel uses each, in which, for $b=1, \ldots,
B$, block $b$ is affected by a flat fading coefficient $h_b\in \mathbb{C}$. Let
$\gamma_b= |h_b|^2$ be the power fading gain and assume that the fading gain
vector $\boldgamma = (\gamma_1, \ldots, \gamma_B)$ is available at both the
transmitter and the receiver. The transmit power is allocated to the blocks
according to the scheme $\mathbf{p}(\boldgamma)=(p_1(\boldgamma), \ldots,
p_B(\boldgamma))$. Then, the complex baseband channel model can be written as
\begin{equation}
\label{eq:channel_model} \mathbf{y}_b = \sqrt{p_b(\boldgamma)}h_b \mathbf{x}_b
+ \mathbf{z}_b, \ \ b= 1, \ldots, B,
\end{equation}
where $\mathbf{y}_b \in \mathbb{C}^L$ is the received signal in block $b$,
$\mathbf{x}_b \in \mathcal{X}^L \subset \mathbb{C}^L$ is the portion of the
codeword being transmitted in block $b$, $\mathcal{X} \subset \mathbb{C}$ is
the signal constellation and $\mathbf{z}_b \in \mathbb{C}^L$ is a noise vector
with independent, identically distributed (i.i.d.) circularly symmetric
Gaussian entries $\sim \mathcal{N}_{\mathbb{C}}(0, 1)$. Assume that the signal
constellation $\mathcal{X}$ is normalized in energy such that
$\sum_{x\in\mathcal{X}}|x|^2 = 2^M$, where $M=\log_2|\mathcal{X}|$. Then, the
instantaneous received SNR at block $b$ is given by $p_b(\boldgamma) \gamma_b$.

We consider block-fading channels where $h_b$ are realizations of a random
variable $H$, whose magnitude is Nakagami-$m$-distributed and has a uniformly
distributed phase\footnote{Due to our perfect transmitter and receiver CSI
assumption, we can assume that the phase has been perfectly compensated for.}.
The fading magnitude has the following probability density function (pdf)
\begin{equation}
f_{|H|}(h)= \frac{2m^m h^{2m-1}}{\Gamma(m)}e^{-m h^2},
\end{equation}
where $\Gamma(a)$ is the Gamma function, $\Gamma(a) =
\int_0^{\infty}t^{a-1}e^{-t}dt$. The coefficients $\gamma_b$ are realizations of the random variable $|H|^2$ whose pdf is given by
\begin{equation}
\label{eq:sq_Naka_dist}
f_{|H|^2}(\gamma)= \begin{cases}
\frac{m^m \gamma^{m-1}}{\Gamma(m)}e^{-m\gamma}, & \gamma \geq 0\\
0, &{\rm otherwise}.
\end{cases}
\end{equation}
The Nakagami-$m$ distribution encompasses many fading distributions of interest. In particular, we obtain Rayleigh fading by letting $m=1$ and Rician fading with parameter $K$ by setting $m=(K+1)^2/(2K+1)$.

\section{Mutual Information and Outage Probability}
\label{se:info_theory}

For any given power fading gain realization $\boldgamma$
and power allocation scheme $\mathbf{p}(\boldgamma)$, the instantaneous
input-output mutual information of the channel is given by
\begin{equation}
I_B(\mathbf{p}(\boldgamma), \boldgamma) = \frac{1}{B}\sum_{b=1}^B \IX(p_b \gamma_b),
\end{equation}
where $\IX(\rho)$ is the input-output mutual information of an AWGN channel
with input constellation ${\mathcal X}$ received SNR $\rho$
\begin{equation}
  \IX(\rho) = M - \mathbb{E}_{X,Z}\left[\log_2\left(\sum_{x'  \in  \mathcal   X}e^{-|\sqrt{\rho}(X-x')+Z|^2+|Z|^2}\right)\right].\nonumber
\end{equation}
Communication is in outage when the instantaneous input-output mutual
information is less than the target rate $R$. For a given power allocation
scheme $\mathbf{p}(\boldgamma)$, the outage probability at communication rate $R$ is given
by  \cite{OzarowShamaiWyner1994,BiglieriProakisShamai1998}
\begin{align}
\label{eq:out_prob}
\Pout(\mathbf{p}(\boldgamma), R) &=\Pr(I_B(\mathbf{p}(\boldgamma), \boldgamma)< R) \nonumber\\
&=\Pr\left(\frac{1}{B} \sum_{b=1}^B \IX(p_b\gamma_b)< R\right).
\end{align}

\section{Short-Term Power Allocation}
\label{se:short_term}

Short-term power allocation schemes are applied for
systems where
the transmit power of each codeword is limited to $BP$. A given short-term
power allocation scheme $\mathbf{p(\boldgamma)}=(p_1, \ldots, p_B)$ must then
satisfy $\sum_{b=1}^B p_b \leq BP$.

\subsection{Optimal Power Allocation}
The optimal short-term power allocation rule $\mathbf{p}^{\rm opt}(\boldgamma)$
is the solution to the outage probability minimization problem
\cite{CaireTariccoBiglieri1999}. Mathematically we express $\mathbf{p}^{\rm
opt}(\boldgamma)$ as
\begin{equation}
\label{eq:opt_short_prob} \mathbf{p}^{\rm opt}(\boldgamma) = \arg
\min_{\substack{{\mathbf p}\in \mathbb{R}^B_+\\ \sum_{b=1}^B p_b = BP} }
\Pout(\mathbf{p(\boldgamma)}, R).
\end{equation}

For short-term power allocation, since the available power can only be
distributed within one codeword, the power allocation scheme that maximizes the
instantaneous mutual information at each channel realization also minimizes the
outage probability. Formally, we have \cite{CaireTariccoBiglieri1999}
\begin{lemma}
\label{le:min_pout_sol} Let $\mathbf{p}^{\rm opt}(\boldgamma)$ be a solution of
the problem
\begin{equation}
\label{eq:prob_max_cap}
\left\{\begin{array}{ll}
{\rm Maximize} & \sum_{b=1}^B \IX(p_b \gamma_b)\\
{\rm Subject \ to}& \sum_{b=1}^B p_b  \leq BP\\
&p_b \geq 0, b=1, \ldots, B.
\end{array}\right.
\end{equation}
Then $\mathbf{p}^{\rm opt}(\boldgamma)$ is a solution of $\eqref{eq:opt_short_prob}$.
\end{lemma}

From \cite{LozanoTulinoVerdu2006}, the solution for problem
\eqref{eq:prob_max_cap} is given by
\begin{equation}
\label{eq:opt_sol_short}
\popt_b(\boldgamma) =
\frac{1}{\gamma_b}\MMSE^{-1}\left(\min\left\{1,
\frac{\nu}{\gamma_b}\right\}\right),
\end{equation}
for $b=1, \dotsc, B$, where $\MMSE(\rho)$ is the minimum mean-squared error
(MMSE) for estimating the input based on the received signal over an AWGN
channel with SNR $\rho$
\begin{equation}
\MMSE(\rho) = 1 - \frac{1}{\pi} \int_{\mathbb{C}} \frac{\left|\sum_{x\in
\mathcal{X}} x e^{-|y-\sqrt{\rho}x|^2}\right|^2}{\sum_{x \in
\mathcal{X}}e^{-|y-\sqrt{\rho}x|^2}} dy
\end{equation}
and $\nu$ is chosen such that the power constraint is satisfied,
\begin{equation}
\sum_{b=1}^B \popt_b = BP.
\end{equation}

The optimal short-term power allocation scheme improves the outage performance
of block-fading systems. However, it does not increase the outage diversity
compared to uniform power allocation, as shown in the following lemma.
\begin{lemma}
\label{le:opt_diver} Consider transmission over the block-fading channel
defined in \eqref{eq:channel_model} with the optimal power allocation scheme
$\mathbf{p}^{\rm opt}(\boldgamma)$ given in \eqref{eq:opt_sol_short}. Assume
input constellation size $|\mathcal{X}|= 2^M$. Further assume that the power
fading gains follow the distribution given in \eqref{eq:sq_Naka_dist}. Then,
for large $P$ and some $ \mathcal{K}_{\rm opt}>0$ the outage probability
behaves as
\begin{equation}
\Pout(\mathbf{\popt}(\boldgamma), R) \doteq \mathcal{K}_{\rm opt}P^{-m d_B(R)},
\end{equation}
where $d_B(R)$ is the Singleton bound given by
\begin{equation}
d_B(R)= 1+ \left\lfloor B\left(1-\frac{R}{M}\right)\right\rfloor
\label{eq:sb}
\end{equation}
\end{lemma}

\subsection{Suboptimal Power Allocation Schemes}
Although the power allocation scheme in \eqref{eq:opt_sol_short} is optimal, it
involves an inverse MMSE function, which may be too complex to implement or
store for specific low-cost systems. Moreover, the MMSE function provides
little insight to the role of each parameter. In this section, we propose power
allocation schemes similar to water-filling that tackle both drawbacks and
perform very close to the optimal solution.

\subsubsection{Truncated water-filling scheme}
\label{se:wlikesol} The complexity of the solution in \eqref{eq:opt_sol_short}
is due to the complex expression of $I_{\mathcal{X}}(\rho)$ in problem
\eqref{eq:prob_max_cap}. Therefore, in order to obtain a simple suboptimal
solution, we find an aproximation for $\IX(\rho)$ in problem
\eqref{eq:prob_max_cap}. The water-filling solution in
\cite{CaireTariccoBiglieri1999} suggests the following approximation of
$\IX(\rho)$
\begin{equation}
\label{eq:target_tw}
I^{\rm tw}(\rho) = \left\{\begin{array}{ll} \log_2(1+\rho), & \rho \leq \beta\\
\log_2(1+\beta), &\rm{otherwise,}
\end{array}\right.
\end{equation}
where $\beta$ is a design parameter to be optimized for best performance.
The resulting suboptimal scheme $\mathbf{\ptw}(\boldgamma)$ is given as a
solution of
\begin{equation}
\label{eq:trunc_wfill_prob}
\left\{ \begin{array}{ll}\rm{Maximize} &\sum_{b=1}^B I^{\rm tw}(p_b \gamma_b)\\
\rm{Subject \ to} & \sum_{b=1}^B p_b \leq BP\\
&p_b \geq 0, \ b=1, \ldots, B.
\end{array}
\right.
\end{equation}

\begin{lemma}
\label{le:trunc_wfill_sol}
A solution to the problem \eqref{eq:trunc_wfill_prob} is given by
\begin{equation}
\label{eq:trunc_wfill_sol}
\ptwfill_b (\boldgamma)= \begin{cases}
\frac{\beta}{\gamma_b}, & {\rm if} \sum_{b=1}^B \frac{\beta}{\gamma_b} \leq
BP\\
\min\left\{\frac{\beta}{\gamma_b}, \left(\eta -
\frac{1}{\gamma_b}\right)_+\right\}, &{\rm otherwise}
\end{cases}
\end{equation}
for $b=1, \ldots, B$, where $\eta$ is chosen such that
\begin{equation}
\sum_{b=1}^B \min\left\{\frac{\beta}{\gamma_b}, \left(\eta -
\frac{1}{\gamma_b}\right)_+\right\} = BP.
\end{equation}
\end{lemma}
Without loss of generality, assume that $\gamma_1 \geq \ldots \geq \gamma_B$,
then, similarly to water-filling, $\eta$ can be determined such that \cite{CaireTariccoBiglieri1999}
\begin{equation}
(k-l)\eta = BP - \sum_{b=1}^l \frac{\beta+1}{\gamma_b}+ \sum_{b=1}^k
\frac{1}{\gamma_b}
\end{equation}
where $k, l$ are integers satisfying $\frac{1}{\gamma_k}< \eta <
\frac{1}{\gamma_{k+1}}$ and $\frac{\beta+1}{\gamma_l}< \eta \leq
\frac{\beta+1}{\gamma_{l+1}}$.

From  Lemma \ref{le:trunc_wfill_sol}, the resulting power allocation scheme is
similar to water-filling, except for the truncation of the allocated power at
$\frac{\beta}{\gamma_b}$. We refer to this scheme as truncated
water-filling.

The outage performance obtained by the truncated water-filling scheme depends
on the choice of the design parameter $\beta$. We now analyze the asymptotic
performance of the outage probability, thus providing some guidance on the
choice of $\beta$.
 \begin{lemma}
\label{le:tw_diver} Consider transmission over the block-fading channel defined
in \eqref{eq:channel_model} with the truncated water-filling power allocation
scheme $\mathbf{p}^{\rm tw}(\boldgamma)$ given in \eqref{eq:trunc_wfill_sol}.
Assume input constellation $\mathcal{X}$ of size $|\mathcal{X}|= 2^M$. Further
assume that the power fading gains follow the distribution given in
\eqref{eq:sq_Naka_dist}. Then, for large $P$, the outage probability
$\Pout(\mathbf{\ptw}(\boldgamma), R)$ is asymptotically upper bounded by
\begin{equation}
\label{eq:trunc_wfill_outprob_bound}
\Pout(\mathbf{\ptw}(\boldgamma), R) \dot{\leq} \mathcal{K}_{\beta} P^{-m
  d_{\beta}(R)},
\end{equation}
where
\begin{equation}
d_{\beta}(R) =
1+\left\lfloor B\left(1-\frac{R}{I_{\mathcal X}(\beta)}\right)\right\rfloor,
\end{equation}
and $I_{\mathcal{X}}(\beta)$ is the input-output mutual
information of an AWGN channel with SNR $\beta$.
\end{lemma}

From the results of Lemmas \ref{le:opt_diver} and \ref{le:tw_diver}, we note that
$\Pout(\mathbf{\ptw}(\boldgamma), R) \geq \Pout(\mathbf{\popt(\boldgamma)},
R)$, and we have that
\begin{equation}
\Pout(\mathbf{\ptw}(\boldgamma), R) \doteq \mathcal{K}_{\rm tw} P^{-md_{\rm
    tw}(R)},
\end{equation}
where $d_{\rm tw}(R)$ satisfies that $d_\beta(R) \leq d_{\rm tw}(R) \leq d_B(R)$.

Therefore, the truncated water-filling scheme is guaranteed to obtain optimal
diversity whenever $d_\beta(R)= d_B(R)$, or equivalently, when
\begin{align}
B\left(1-\frac{R}{I_{\mathcal X}(\beta)}\right) &\geq
\left\lfloor B\left(1-\frac{R}{M}\right)\right\rfloor\\
I_{\mathcal X}(\beta) &\geq \frac{BR}{B- \left\lfloor B
  \left(1-\frac{R}{M}\right)\right\rfloor}
  \label{eq:boundbeta}
\end{align}
which implies that
\[
\beta \geq I_{\mathcal X}^{-1} \left(\frac{BR}{B- \left\lfloor B
\left(1-\frac{R}{M}\right)\right\rfloor}\right) \triangleq \beta_R.
\]
Therefore, the truncated water-filling power allocation scheme
\eqref{eq:trunc_wfill_sol} becomes the classical water-filling algorithm for
Gaussian inputs, and provides optimal outage diversity at any transmission rate
by letting $\beta \rightarrow \infty$. For any rate $R$ that is not at the
discontinuity points of the Singleton bound, i.e. $R$ such that
$B\left(1-\frac{R}{M}\right)$ is not an integer, we can always design a
truncated water-filling scheme that obtains optimal diversity by choosing
$\beta \geq \beta_R$.

\begin{figure}
\begin{center}
\includegraphics[width  =0.9\columnwidth]
        {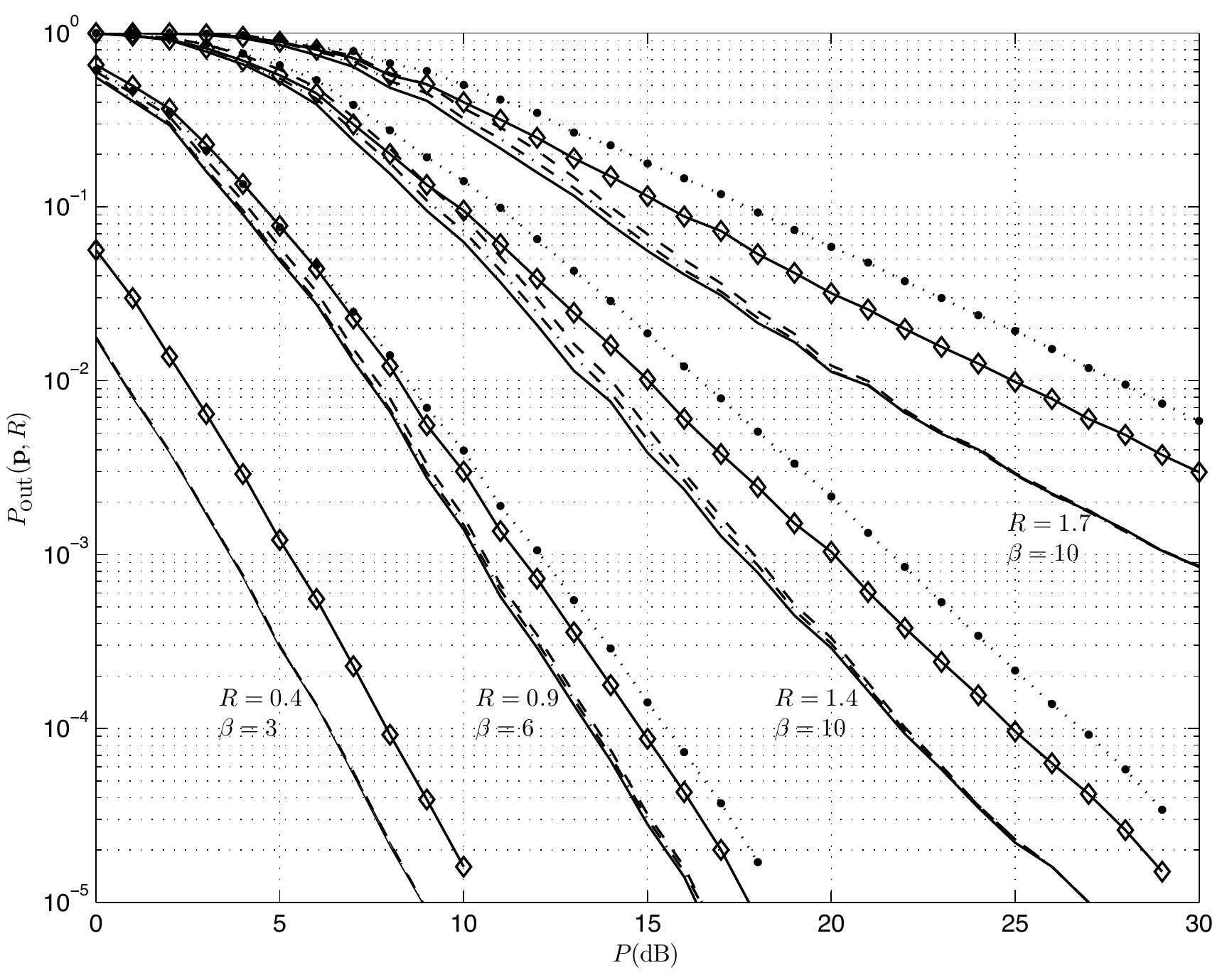}
\vspace{-3mm}\caption{Outage performance of various short-term power allocation schemes for
QPSK-input block-fading channels with $B=4$ and Rayleigh fading. The solid-line
represents optimal scheme; the solid line with $\diamond$ represents uniform
power allocation; the dashed line and dashed-dotted line represent truncated
water-filling and its corresponding refinement, respectively; the dotted line represents classical
water-filling scheme.} \label{fig:QPSK4blocksR0p5_0p9_1p4_1p7_beta10}
\end{center}
\vspace{-5mm}
\end{figure}

With the results above, we choose $\beta$ as follows. For a
transmission rate $R$ that is not a discontinuity point of the Singleton
bound, we perform a simulation to compute the outage probability at
rate $R$ obtained by truncated water-filling with various $\beta \geq \beta_R$
and pick the $\beta$ that gives the best outage performance. The dashed line in
Figure \ref{fig:QPSK4blocksR0p5_0p9_1p4_1p7_beta10} illustrates the performance
of the obtained schemes for block-fading channels with $B=4$, QPSK input under
Rayleigh fading. At all rates of interest, the truncated water-filling schemes
perform very close to the optimal scheme (solid line), especially at high SNR.

For rates at the discontinuous points of the Singleton bound, especially when
operating at high SNR, $\beta$ needs to be relatively large in order to maintain
diversity. However, large $\beta$ increases the gap between $I^{\rm tw}(\rho)$
and $\IX(\rho)$, thus degrades the performance of the truncated water-filling
scheme. For $\beta=15$, the gap is illustrated by the dashed lines in Figure
\ref{fig:ref_trunc_wf_beta15}. In the extreme case where $\beta \to \infty$,
the truncated water-filling turns into the water-filling scheme, which exhibits
a significant loss in outage performance as illustrated
by the dotted lines in Figure \ref{fig:QPSK4blocksR0p5_0p9_1p4_1p7_beta10}. To
reduce this drawback, we propose a better approximation to $\IX(\rho)$, which
leads to a refinement to the truncated water-filling scheme in the next
section.

\subsubsection{Refined truncated water-filling schemes}
\label{se:refinement}
 To obtain better approximation to the optimal power allocation scheme, we need
 a more accurate approximation to $\IX(\rho)$ in \eqref{eq:prob_max_cap}. We propose
 the following approximation

 \begin{equation}
 \label{eq:ref_target}
 I^{\rm ref}(\rho) = \left\{\begin{array}{ll}
 \log_2(1+\rho), &\rho \leq \alpha\\
 \kappa \log_2(\rho)+ a, &\alpha < \rho \leq \beta \\
 \kappa \log_2(\beta)+ a, &{\rm otherwise,}
 \end{array}\right.
 \end{equation}
where $\kappa$ and $a$ are chosen such that in dB scale, $\kappa
\log_2(\rho)+ a$ is a tangent to $\IX(\rho)$ at a predetermined point $\rho_0$.
Therefore $\alpha$ is chosen such that $\kappa\log_2(\alpha)+a = \log_2(1+\alpha)$, and
$\beta$ is a design parameter. For QPSK input and $\rho_0 = 3$, we have
$\kappa= 0.3528, a= 1.1327, \alpha = 1.585$.

 The optimization problem \eqref{eq:prob_max_cap} is approximated by
 \begin{equation}
\label{eq:ref_prob}
 \left\{\begin{array}{ll}
 {\rm Maximize} &\sum_{b=1}^B  I^{\rm ref}(p_b \gamma_b)\\
 {\rm Subject \ to} & \sum_{b=1}^B p_b \leq BP\\
 &p_b \geq 0, b=1, \ldots, B.
 \end{array}\right.
 \end{equation}
The refined truncated
water-filling scheme $\mathbf{\pref}(\boldgamma)$ is given by the following
lemma.
\begin{lemma}
\label{le:ref_sol} A solution to problem \eqref{eq:ref_prob} is
\begin{equation}
\pref_b = \frac{\beta}{\gamma_b}, \,\,\,b=1, \ldots, B
\end{equation}
if $\sum_{b=1}^B \frac{\beta}{\gamma_b} < BP$, and otherwise, for $b=1, \ldots,
B$,
\begin{equation}
\pref_b = \begin{cases}
\frac{\beta}{\gamma_b}, &\eta \geq \frac{\beta}{\kappa \gamma_b}\\
\kappa \eta, &\frac{\alpha}{\kappa \gamma_b} \leq \eta <
\frac{\beta}{\kappa  \gamma_b}\\
\frac{\alpha}{\gamma_b}, &\frac{\alpha+1}{\gamma_b}\leq \eta <
\frac{\alpha}{\kappa\gamma_b}\\
\eta-\frac{1}{\gamma_b}, &\frac{1}{\gamma_b} \leq \eta <
\frac{\alpha+1}{\gamma_b}\\
0, &{\rm otherwise,}
\end{cases}
\end{equation}
where $\eta$ is chosen such that
\begin{equation}
\sum_{b=1}^B \pref_b = BP.
\end{equation}
\end{lemma}

\begin{figure}
\begin{center}
\includegraphics[width =0.9\columnwidth]{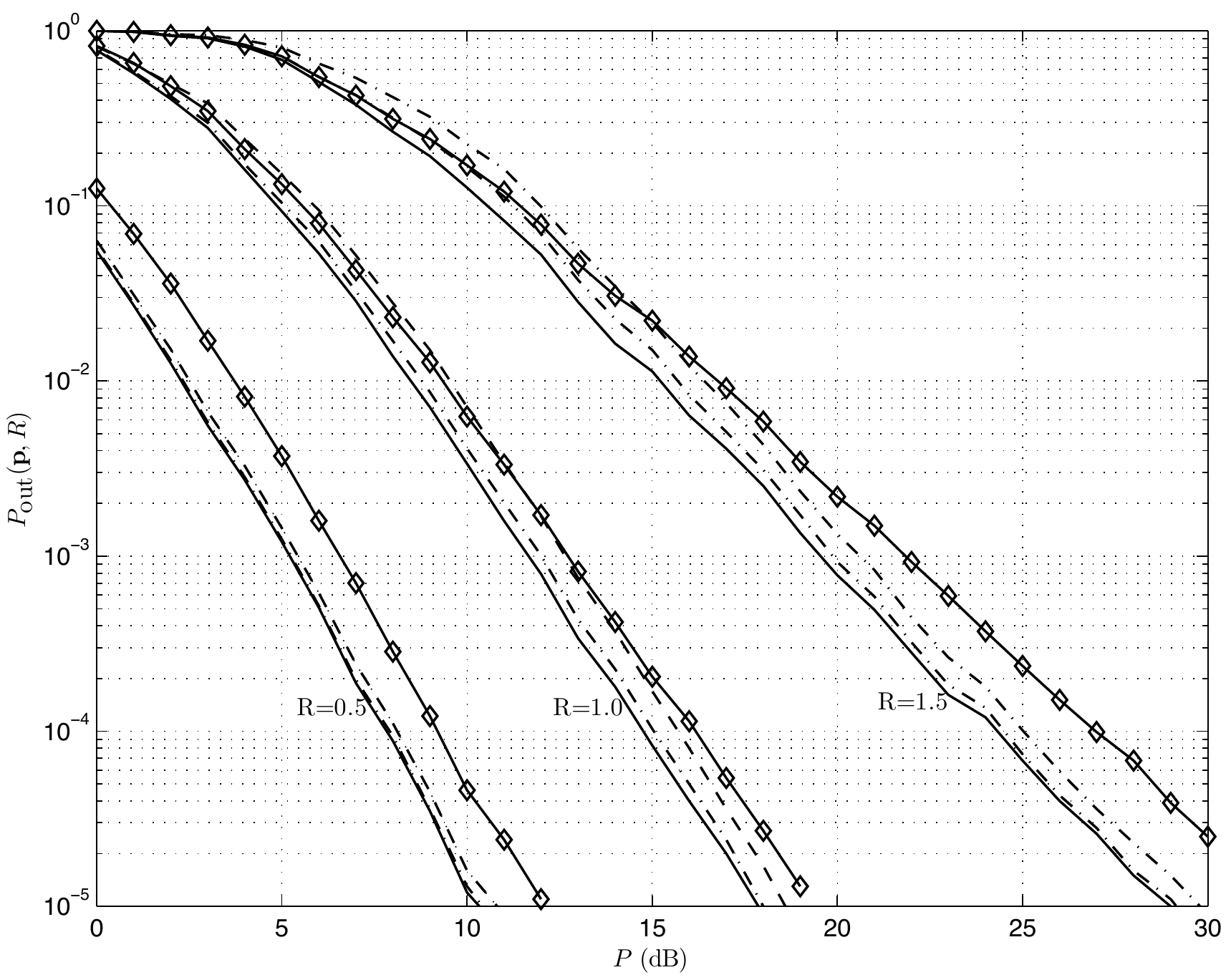}
\vspace{-3mm}\caption{Outage performance of various short-term power allocation schemes for
QPSK-input block-fading  channels with $B=4$ and Rayleigh fading. The solid-line
  represents optimal scheme; the solid line with $\diamond$ represents uniform
  power allocation; the dashed line and dashed-dotted line correspondingly
  represent truncated water-filling and its refinement with $\beta = 15$.}
\label{fig:ref_trunc_wf_beta15}
\end{center}
\vspace{-5mm}
\end{figure}

The refined truncated water-filling scheme provides significant gain over the
truncated water-filling scheme, especially when the transmission rate requires
relatively large $\beta$ to maintain the outage diversity. The dashed-dotted
lines in Figure \ref{fig:ref_trunc_wf_beta15} show the outage performance of
the refined truncated water-filling scheme for block-fading channels with
$B=4$, QPSK input under Rayleigh fading. The refined truncated water-filling
scheme performs very close to the optimal case even at the rates where the
Singleton bound is discontinuous, i.e. rates $R= 0.5, 1.0, 1.5$. The
performance gains of the refined scheme over the truncated water-filling scheme
at other rates are also illustrated by the dashed-dotted lines in Figure
\ref{fig:QPSK4blocksR0p5_0p9_1p4_1p7_beta10}.

\section{Long-Term Power Allocation}
\label{se:long_term}

We consider systems with long-term power constraints, in
which the expectation of the power allocated to each block (over infinitely
many codewords) does not exceed $P$. This problem has been investigated in
\cite{CaireTariccoBiglieri1999} for block-fading channels with Gaussian
inputs. In this section, we obtain similar results for channels with discrete
inputs, and propose suboptimal schemes that reduce the complexity of the
algorithm.

\subsection{Optimal Long-Term Power Allocation}
Following \cite{CaireTariccoBiglieri1999}, the problem can be formulated as
\begin{equation}
\label{eq:long_prob}
\left\{\begin{array}{ll} {\rm Minimize }& \Pr(I_B(\plt(\boldgamma),
\boldgamma)< R)\\
{\rm Subject \ to}& \expectation{\mean{\plt(\boldgamma)}}{} \leq P,\\
\end{array}\right.
\end{equation}
where $\mean{\mathbf{p}} = \frac{1}{B}\sum_{b=1}^B p_b$.

The following theorem shows that the structure of the optimal long-term
solution $\pltopt(\boldgamma)$ of \cite{CaireTariccoBiglieri1999} for Gaussian
inputs is generalized to the discrete-input case.
\begin{theorem}
\label{theo:opt_longterm}
Problem \eqref{eq:long_prob} is solved by $\pltopt(\boldgamma)$ given by
\begin{equation}
\label{eq:long_term}
\pltopt(\boldgamma) = \left\{\begin{array}{ll}
\boldlp^{\rm opt}(\boldgamma), &{\rm if\ }\boldgamma \in \mathcal{R}(\sstar)\\
0, &{\rm if\ } \boldgamma \notin \overline{\mathcal{R}}(\sstar),
\end{array}\right.
\end{equation}
while if $\boldgamma \in \overline{\mathcal{R}}(\sstar) \setminus
\mathcal{R}(\sstar)$  then $\pltopt(\boldgamma)= \boldlp(\boldgamma)$ with
probability $\wstar$ and $\pltopt(\boldgamma)= 0$ with probability $1-
\wstar$, where $\boldlp(\boldgamma)$ is the solution of the following
optimization problem
\begin{equation}
\label{eq:opt_min_pow_prob}
\left\{\begin{array}{ll}
{\rm Minimize }& \mean{\boldlp}\\
{\rm Subject \ to }& \sum_{b=1}^B \IX(\lp_b \gamma_b) \geq BR\\
&\lp_b \geq 0, b=1, \ldots, B,
\end{array}\right.
\end{equation}
and $\mathcal{R}(s), \overline{\mathcal R}(s), \sstar, \wstar$
are defined as follows
\begin{align}
\label{eq:def_Rs}
{\mathcal R}(s)& = \{\boldgamma \in \mathbb{R}_+^B: \mean{\boldlp^{\rm opt}(\boldgamma)}
< s\}\\
\overline{\mathcal R}(s)&=\{\boldgamma \in \mathbb{R}_+^B:
\mean{\boldlp^{\rm opt}(\boldgamma)} \leq s\}\\
\sstar &= \sup\{s: P(s) < P\}\\
\wstar &= \frac{P- P(\sstar)}{\overline{P}(\sstar)- P(\sstar)}
\end{align}
where\footnote{For simplicity, for a random variable $\xi$ with pdf
$f_{\xi}(\xi)$, we denote $\mathbb{E}_{\xi \in \mathcal{A}}[f(\xi)] \triangleq
\int_{\xi \in \mathcal A}f_{\xi}(\xi)d\xi$.}
\begin{align}
P(s)&= \mathbb{E}_{\boldgamma \in \mathcal{R}(s)}\left[\mean{\boldlp^{\rm
      opt}(\boldgamma)}\right]\\
\label{eq:def_ovlinePs}
\overline{P}(s)&= \mathbb{E}_{\boldgamma \in
      \overline{\mathcal{R}}(s)}\left[\mean{\boldlp^{\rm
      opt}(\boldgamma)}\right]
\end{align}
and $\boldlp^{\rm opt}(\boldgamma)$ is the solution of \eqref{eq:opt_min_pow_prob} given by
\begin{equation}
\lp^{\rm opt}_b = \left\{\begin{array}{ll}
\frac{1}{\gamma_b}\MMSE^{-1}\left(\frac{1}{\eta \gamma_b}\right), & \eta \geq
\frac{1}{\gamma_b}\\
0, &{\rm otherwise}
\end{array}\right.
\end{equation}
for $b=1, \ldots, B$, where $\eta$ is chosen such that
\begin{equation}
\sum_{b=1; \gamma_b \geq \frac{1}{\eta}}^B
\IX\left(\MMSE^{-1}\left(\frac{1}{\eta \gamma_b}\right)\right) = BR.
\end{equation}
\end{theorem}

As in the Gaussian input case \cite{CaireTariccoBiglieri1999}, the optimal
power allocation scheme either transmits with the minimum power that enables
transmission at the target rate, or turns off transmission (allocating zero
power) when the channel realization is bad.
Therefore, there is no power wastage on outage events.

The solid lines in Figure \ref{fig:long_schemes} illustrates the outage
performance of optimal long-term power allocation schemes for transmission over
4-block block-fading channels with QPSK-input and Rayleigh fading. The
simulation results suggest that for communication rates where $d_B(R)> 1$, zero
outage probability can be obtained with finite power. This agrees to the
results obtained for block-fading channels with Gaussian inputs
\cite{CaireTariccoBiglieri1999}, where only for $B>1$ zero outage was possible.

\subsection{Suboptimal Long-Term Power Allocation}

In the optimal long-term power allocation scheme $\plt^{\rm opt}(\boldgamma)$
given in Theorem \ref{theo:opt_longterm}, $\wstar, \sstar$ can be evaluated
offline for any fading distribution. Therefore, given an allocation scheme
$\boldlp^{\rm opt}(\boldgamma)$, the complexity required to evaluate $\plt^{\rm
opt}(\boldgamma)$ is low. Thus, the complexity of the long-term power
allocation scheme is mainly due to the complexity of evaluating $\boldlp^{\rm
  opt}(\boldgamma)$, which requires the evaluation or storage of $\MMSE(\rho)$ and
$\IX(\rho)$. In this section, we propose suboptimal long-term power allocation
schemes by replacing $\boldlp^{\rm opt}(\boldgamma)$ with simpler power
allocation algorithms.

  A long-term power allocation scheme $\plt(\boldgamma)$ corresponding to an
  arbitrary $\boldlp(\boldgamma)$ is obtained by replacing $\boldlp^{\rm
  opt}(\boldgamma)$ in \eqref{eq:long_term},
  \eqref{eq:def_Rs}--\eqref{eq:def_ovlinePs} with $\boldlp(\boldgamma)$.
From \eqref{eq:long_term}, \eqref{eq:def_Rs}--\eqref{eq:def_ovlinePs}, the
long-term power allocation scheme $\plt(\boldgamma)$ satisfies
\begin{align}
\expectation{\mean{\plt(\boldgamma)}}{}= &\expectation{\mean{\plt(\boldgamma)}}{\boldgamma \in \mathcal{R}(\sstar)}\\
&+\wstar\expectation{\plt(\boldgamma)}{\boldgamma \in
\overline{\mathcal{R}}(\sstar)\setminus \mathcal{R}(\sstar)}\\
= &P(\sstar)+ \wstar\left(\overline{P}(\sstar)-P(\sstar)\right)=P
\end{align}
Therefore, a long-term power allocation schemes corresponding to an arbitrary
$\boldlp(\boldgamma)$ is suboptimal with respect to $\plt^{\rm
  opt}(\boldgamma)$. Following the transmission strategy in the optimal scheme,
we consider the power allocation schemes $\boldlp(\boldgamma)$ that satisfy the
rate constraint $I_B(\boldlp(\boldgamma), \boldgamma)\geq R$ to avoid wasting
power on outage events. These schemes are suboptimal solutions of problem
\eqref{eq:opt_min_pow_prob}. Based on the short-term schemes, two simple rules are discussed in the next
subsections.

\subsubsection{Long-term truncated water-filling scheme}
\label{se:tw_long}
 Similar to the short-term truncated water-filling scheme,
we consider approximating $\IX(\rho)$ in \eqref{eq:opt_min_pow_prob} by $I^{\rm
tw}(\rho)$ in \eqref{eq:target_tw}, which results in the following problem
\begin{equation}
\label{eq:tw_min_pow_prob}
\left\{\begin{array}{ll}
{\rm Minimize} &\mean{\boldlp(\gamma)}\\
{\rm Subject \ to \ }&\sum_{b=1}^B I^{\rm tw}(\lp_b \gamma_b)\geq BR\\
&\lp_b \geq 0, b=1, \ldots, B
\end{array}\right.
\end{equation}
The solution of \eqref{eq:tw_min_pow_prob} is given by
\begin{equation}
\label{eq:ptw_org} \lp_b = \min\left\{\frac{\beta}{\gamma_b}, \left(\eta -
  \frac{1}{\gamma_b}\right)_+ \right\}, b=1, \ldots, B,
\end{equation}
where $\eta$ is chosen such that
\begin{equation}
\sum_{b=1}^B \log_2(1+\lp_b \gamma_b) = BR.
\end{equation}

Note that since $I^{\rm tw}(\rho)$ upperbounds $\IX(\rho)$,
$\boldlp(\boldgamma)$ does not satisfy the rate constraint
$I_B(\boldlp(\boldgamma), \boldgamma)\geq R$. By adjusting $\eta$, we can
obtain a suboptimal $\boldlp^{\rm tw}(\boldgamma)$ of $\boldlp^{\rm
opt}(\boldgamma)$ as follows
\begin{equation}
\label{eq:lp_tw} \lp^{\rm tw}_b = \min\left\{\frac{\beta}{\gamma_b}, \left(\eta
-
  \frac{1}{\gamma_b}\right)_+ \right\}, b=1, \ldots, B,
\end{equation}
where $\eta$ is chosen such that
\begin{equation}
\sum_{b=1}^B \IX(\lp^{\rm tw}_b \gamma_b) = BR.
\end{equation}
Using this scheme, we obtain a power allocation $\plt^{\rm tw}(\boldgamma)$,
which is the long-term power allocation scheme corresponding to the suboptimal
$\boldlp^{\rm tw}(\boldgamma)$ of $\boldlp^{\rm opt}(\boldgamma)$. The
performance of the scheme is illustrated by the dashed lines in Figure
\ref{fig:long_schemes}.

\subsubsection{Refinement of the long-term truncated water-filling}
\label{se:ref_long} 
\begin{figure}
\begin{center}
\includegraphics[width =0.9\columnwidth]{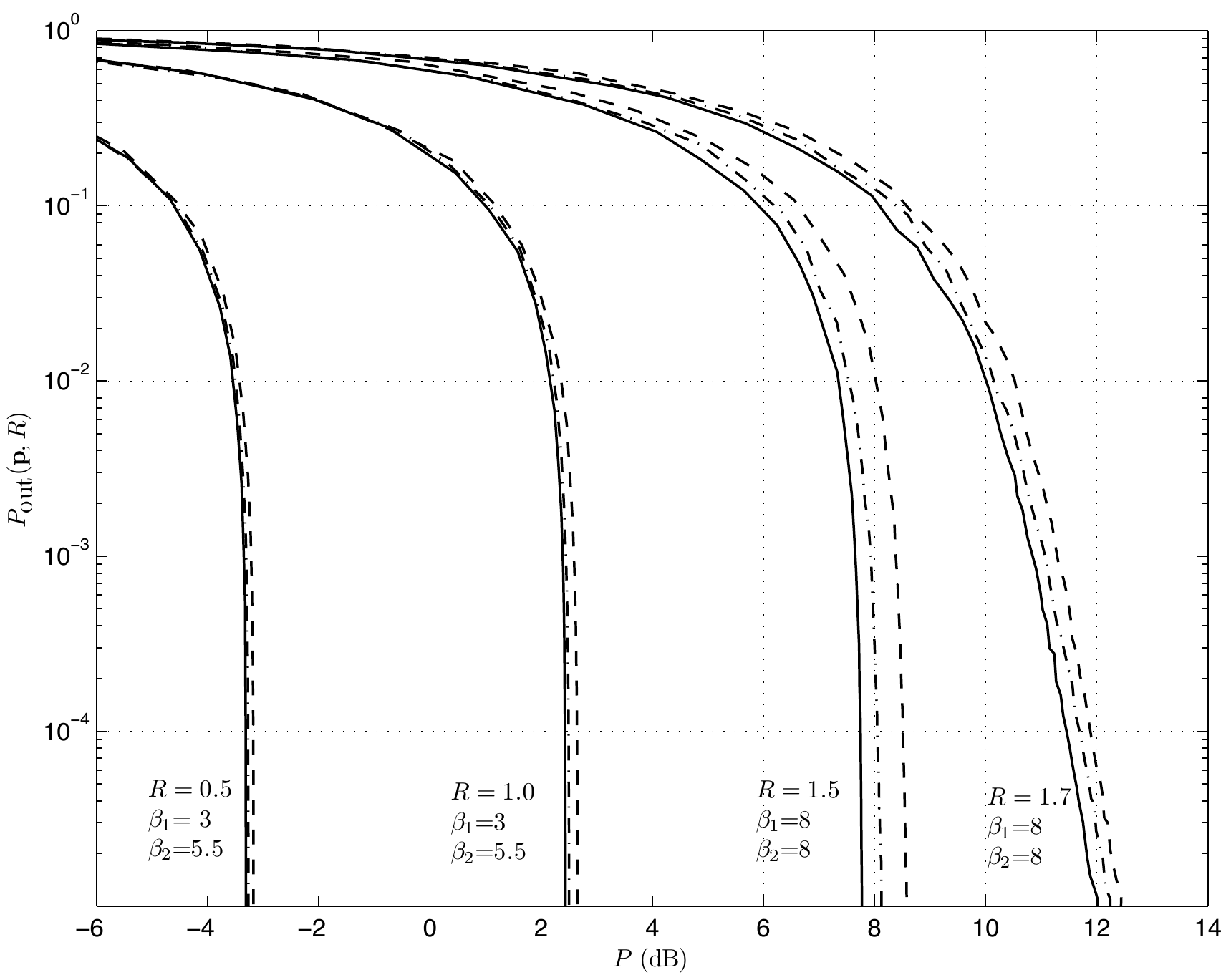}
\vspace{-5mm}\caption{Outage performance of various long-term power allocation schemes for
QPSK-input
  4-block block-fading  channels under Rayleigh fading. The solid-line
  represents optimal scheme; the dashed line and dashed-dotted line correspondingly
  represent long-term truncated water-filling ($\mathbf{p}_{\rm lt}^{\rm tw}(\boldgamma)$ with $\beta_1$)
  and its refinement ($\mathbf{p}_{\rm lt}^{\rm ref}(\boldgamma)$ with $\beta_2$).}
\label{fig:long_schemes}
\end{center}
\vspace{-3mm}
\end{figure}

In order to improve the performance of the suboptimal
scheme, we approximate $\IX(\rho)$ by $I^{\rm ref}(\rho)$ given in
\eqref{eq:ref_target}. Replacing $\IX(\rho)$ in \eqref{eq:opt_min_pow_prob} by
$I^{\rm ref}(\rho)$, we have the following problem
\begin{equation}
\label{eq:ref_min_pow_prob} \left\{\begin{array}{ll}
{\rm Minimize} &\mean{\boldlp(\gamma)}\\
{\rm Subject \ to \ }&\sum_{b=1}^B I^{\rm ref}(\lp_b \gamma_b)\geq BR\\
&\lp_b \geq 0, b=1, \ldots, B
\end{array}\right.
\end{equation}
Following the same steps as in Section \ref{se:tw_long}, the suboptimal
$\boldlp^{\rm ref}(\boldgamma)$ of $\boldlp^{\rm opt}(\boldgamma)$ is given as
\begin{equation}
\lp^{\rm ref}_b = \left\{\begin{array}{ll} \frac{\beta}{\gamma_b}, &\eta \geq
\frac{\beta}{\kappa \gamma_b}\\
  \kappa \eta, &\frac{\alpha}{\kappa \gamma_b} \leq \eta \leq \frac{\beta}{\kappa
  \gamma_b}\\
  \frac{\alpha}{\gamma_b}, &\frac{\alpha+1}{\gamma_b}\leq \eta \leq
  \frac{\alpha}{\kappa\gamma_b}\\
  \eta-\frac{1}{\gamma_b}, &\frac{1}{\gamma_b}\leq \eta \leq
  \frac{\alpha+1}{\gamma_b}\\
  0, &{\rm otherwise,}
\end{array}\right.
\end{equation}
where $\eta$ is chosen such that
\begin{equation}
\sum_{b=1}^B \IX(\lp^{\rm ref}_b \gamma_b)= BR.
\end{equation}
The performance of the long-term power allocation corresponding to
$\boldlp^{\rm ref}(\boldgamma)$, $\plt^{\rm ref}(\boldgamma)$, is illustrated
by the dashed-dotted lines in Figure \ref{fig:long_schemes}.

\subsubsection{Approximation of $\IX(\rho)$}
The suboptimal schemes in the previous sections perform close to optimality,
and are simpler than the optimal scheme. However, the suboptimal schemes still
require the implementation or storage of $\IX(\rho)$ to compute $\eta$. This can be avoided by using
approximations of $\IX(\rho)$. Let $\tilde{I}_{\mathcal X}(\rho)$ be an
approximation of $\IX(\rho)$ and the rate error $\Delta R = \max_{\rho} \{\tilde{I}_{\mathcal X}(\rho)-\IX(\rho)\}$.
Then, for a suboptimal scheme $\boldlp(\boldgamma)$, $\eta$ chosen such that
\begin{equation}
\sum_{b=1}^B \tilde{I}_{\mathcal X}(\lp_b \gamma_b) = B(R+\Delta R)
\end{equation}
satisfies the rate constraint since 
\begin{align}
\sum_{b=1}^B \IX(\lp_b \gamma_b) \geq \sum_{b=1}^B \tilde{I}_{\mathcal X}(\lp_b
\gamma_b)- B \Delta R = BR.
\end{align}

Following \cite{BrannstromRasmussenGrant2005}, we use the
approximation for $\IX(\rho)$
\begin{equation}
\tilde{I}_{\mathcal X}(\rho) = M\left(1-e^{-c_1 \rho^{c_2}}\right)^{c_3}.
\end{equation}
 For channels with QPSK input, using numerical optimization to minimize
the mean squared error between $\IX(\rho)$ and $\tilde{I}_{\mathcal X}(\rho)$,
we obtain $c_1=0.77, c_2=0.87, c_3=1.16$ and $\Delta R= 0.0033$. Using this
approximation to evaluate $\eta$ in subsections \ref{se:tw_long} and
\ref{se:ref_long}, we arrive at much less computationally demanding power
allocation schemes with little loss in performance.

\begin{figure}
\begin{center}
\includegraphics[width =0.9\columnwidth]{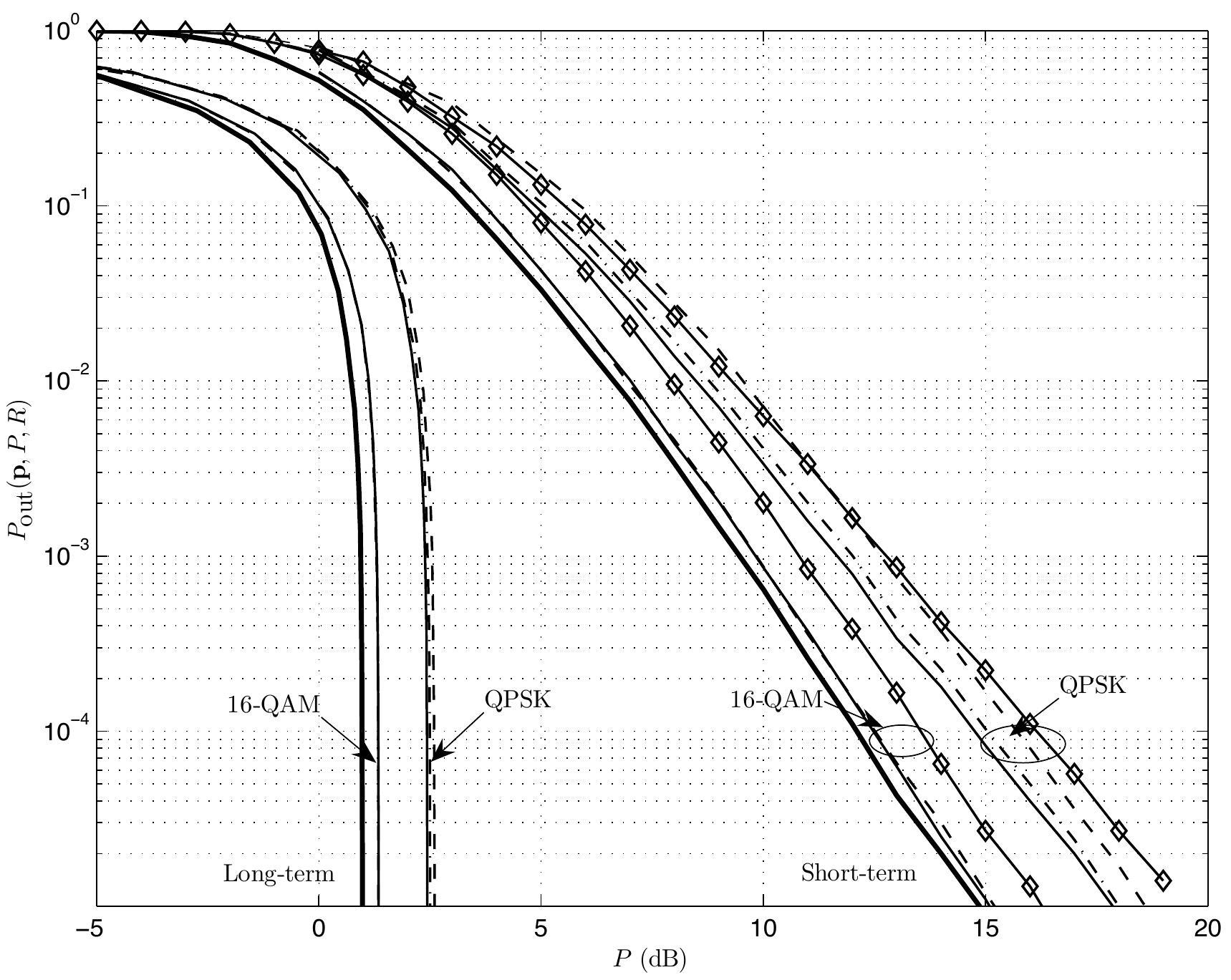}
\vspace{-5mm}\caption{Outage performance of short- and long-term power allocation schemes in a block-fading channel with $B=4$, $R=1$ and Rayleigh fading. Diamonds correspond to uniform allocation, thick solid lines correspond to Gaussian input, thin solid-lines
  represent optimal schemes; the dashed and dashed-dotted lines
  represent long-term truncated water-filling ($\mathbf{p}_{\rm lt}^{\rm tw}(\boldgamma)$ with $\beta_1=3$)
  and its corresponding refinement ($\mathbf{p}_{\rm lt}^{\rm ref}(\boldgamma)$ with $\beta_2=5.5$), respectively.}
\label{fig:comp_short_long}
\end{center}
\vspace{-5mm}
\end{figure}

We finally illustrate in Figure \ref{fig:comp_short_long} the significant gains
achievable by the long-term schemes when compared to short-term. As remarked in
\cite{CaireTariccoBiglieri1999}, remarkable gains are possible with Gaussian
inputs (11dB at $10^{-4}$). As shown in the figure, similar gains (12dB at
$10^{-4}$) are also achievable by discrete inputs. Note that, due to the
Singleton bound, the slope of the discrete-input short-term curves is not as
steep as the slope of the corresponding Gaussian input curve.

\section{Conclusion}
\label{se:conclude}

We considered power allocation schemes for discrete-input block-fading channels
with transmitter and receiver CSI under short- and long-term power constraints.
We have studied optimal and low-complexity sub-optimal schemes, and have
illustrated the corresponding performances, showing that minimal loss is
incurred when using the sub-optimal schemes.

\bibliographystyle{IEEEtran}
\bibliography{bib_database}

\begin{thebibliography}{10}
\providecommand{\url}[1]{#1}
\csname url@rmstyle\endcsname
\providecommand{\newblock}{\relax}
\providecommand{\bibinfo}[2]{#2}
\providecommand\BIBentrySTDinterwordspacing{\spaceskip=0pt\relax}
\providecommand\BIBentryALTinterwordstretchfactor{4}
\providecommand\BIBentryALTinterwordspacing{\spaceskip=\fontdimen2\font plus
\BIBentryALTinterwordstretchfactor\fontdimen3\font minus
  \fontdimen4\font\relax}
\providecommand\BIBforeignlanguage[2]{{%
\expandafter\ifx\csname l@#1\endcsname\relax
\typeout{** WARNING: IEEEtran.bst: No hyphenation pattern has been}%
\typeout{** loaded for the language `#1'. Using the pattern for}%
\typeout{** the default language instead.}%
\else
\language=\csname l@#1\endcsname
\fi
#2}}

\bibitem{OzarowShamaiWyner1994}
L.~H. Ozarow, S.~Shamai, and A.~D. Wyner, ``Information theoretic
  considerations for cellular mobile radio,'' \emph{IEEE Trans. Veh. Tech.},
  vol.~43, no.~2, pp. 359--378, May 1994.

\bibitem{BiglieriProakisShamai1998}
E.~Biglieri, J.~Proakis, and S.~Shamai, ``Fading channels: Informatic-theoretic
  and communications aspects,'' \emph{IEEE Trans. Inf. Theory}, vol.~44, no.~6,
  pp. 2619--2692, Oct. 1998.

\bibitem{CoverThomas2006}
T.~M. Cover and J.~A. Thomas, \emph{Elements of Information Theory},
  2nd~ed.\hskip 1em plus 0.5em minus 0.4em\relax John Wiley and Sons, 2006.

\bibitem{CaireTariccoBiglieri1999}
G.~Caire, G.~Taricco, and E.~Biglieri, ``Optimal power control over fading
  channels,'' \emph{IEEE Trans. Inf. Theory}, vol.~45, no.~5, pp. 1468--1489,
  Jul. 2001.

\bibitem{LozanoTulinoVerdu2006}
A.~Lozano, A.~M. Tulino, and S.~Verd\'{u}, ``Opitmum power allocation for
  parallel {G}aussian channels with arbitrary input distributions,'' \emph{IEEE
  Trans. Inf. Theory}, vol.~52, no.~7, pp. 3033--3051, Jul. 2006.

\bibitem{KnoppHumblet2000}
R.~Knopp and P.~A. Humblet, ``On coding for block fading channels,'' \emph{IEEE
  Trans. Inf. Theory}, vol.~46, no.~1, pp. 189--205, Jan. 2000.

\bibitem{MalkamakiLeib1999}
E.~Malkam\"{a}ki and H.~Leib, ``Coded diversity on block-fading channels,''
  \emph{IEEE Trans. Inf. Theory}, vol.~45, no.~2, pp. 771--781, Mar. 1999.

\bibitem{FabregasCaire2006}
A.~{Guill\'{e}n i F\`{a}bregas} and G.~Caire, ``Coded modulation in the
  block-fading channel: Coding theorems and code construction,'' \emph{IEEE
  Trans. Inf. Theory}, vol.~52, no.~1, pp. 91--114, Jan. 2006.

\bibitem{NguyenGuillenRasmussen2007}
K.~D. Nguyen, A.~{Guill\'{e}n i F\`{a}bregas}, and L.~K. Rasmussen, ``Power
  allocation for discrete-input delay-limited fading channels,''
  \emph{submitted to IEEE Trans. Inf. Theory. Available at arXiv:0706.2033}.

\bibitem{BrannstromRasmussenGrant2005}
F.~Br\"{a}nnstr\"{o}m, L.~K. Rasmussen, and A.~J. Grant, ``Convergence analysis
  and optimal scheduling for multiple concatenated codes,'' \emph{IEEE Trans.
  Inf. Theory}, vol.~51, no.~9, pp. 3354--3364, Sep. 2005.

\end{thebibliography}

\end{document}